\documentclass[aps,prl,showpacs,twocolumn,groupedaddress]{revtex4}

\usepackage{graphics}
\usepackage{epsfig}

\newcommand{\bz}{\bf\hat z}
\def\R{{\bf R}}

\def\S{{\bf S}}
\def\O{{\bf \Omega}}

\def\bu{\mbox{\boldmath  $u$}}

\begin{document}

\title{Asymmetric, helical and mirror-symmetric travelling waves in
pipe flow}

\author{Chris Pringle}   
\email{Chris.Pringle@bris.ac.uk}
\author{Rich R. Kerswell}
\email{R.R.Kerswell@bris.ac.uk}

\affiliation{Department of Mathematics, University of Bristol, University Walk,
   Bristol BS8 1TW, United Kingdom}

\date{\today}

\begin{abstract}
  
  New families of three-dimensional nonlinear travelling waves are
  discovered in pipe flow. In contrast to known waves (Faisst \&
  Eckhardt {\em Phys. Rev. Lett.}  {\bf 91}, 224502 (2003), Wedin \&
  Kerswell, {\em J. Fluid Mech.} {\bf 508}, 333 (2004)), they possess
  no rotational symmetry and exist at much lower Reynolds numbers.
  Particularly striking is an `asymmetric mode' which has one slow
  streak sandwiched between two fast streaks located preferentially to
  one side of the pipe. This family originates in a pitchfork
  bifurcation from a mirror-symmetric travelling wave which can be
  traced down to a Reynolds number of 773.  Helical and non-helical
  rotating waves are also found emphasizing the richness of phase
  space even at these very low Reynolds numbers. The delay in Reynolds
  number from when the laminar state ceases to be a global
  attractor to turbulent transition is then even larger than
  previously thought.

\end{abstract}


\maketitle

% introduction

Wall-bounded shear flows are of tremendous practical importance yet
their transition to turbulence is still poorly understood. The oldest
and most famous example is the stability of flow along a straight pipe
of circular cross-section first studied over 120 years ago
\cite{reynolds83}.  A steady, unidirectional, laminar solution -
Hagen-Poiseuille flow \cite{hagen39,pois40} - always exists but is
only realised experimentally for lower flow rates (measured by the
Reynolds number $Re=UD/\nu$, where $U$ is the mean axial flow speed,
$D$ is the pipe diameter and $\nu$ is the fluid's kinematic
viscosity). At higher $Re$, the fluid selects a state which is
immediately spatially and temporally complex rather than adopting a
sequence of intermediate states of gradually decreasing symmetry.  The
exact transition Reynolds number $Re_t$ depends sensitively on the
shape and amplitude of the disturbance present and therefore varies
across experiments with quoted values typically ranging from $2300$
down to a more recent estimate of 1750
(\cite{binnie47,lingren58,leite59,wygnanski73,darbyshire95,hof03,peixinho05,peixinho06,willis07}).
A new direction in rationalising this abrupt transition revolves
around identifying alternative solutions (beyond the laminar state) to
the governing Navier-Stokes equations. These have only recently be
found in the form of travelling waves (TWs) \cite[]{faisst03,wedin04}
and all seem to be saddle points in phase space. They appear though
saddle node bifurcations with the lowest found at $Re=1251$. The delay
before transition occurs ($Re_t \geq 1750$) is attributed to the need
for phase space to become sufficiently complicated (through the
entanglement of stable and unstable manifolds of an increasing number
of saddle points) to support turbulent trajectories.

In this Letter, we present four new families of `asymmetric',
`mirror-symmetric', helical and non-helical rotating TWs which have
different structure to known solutions and exist at lower Reynolds
numbers. The asymmetric family, which have one slow streak sandwiched
between two fast streaks located preferentially to one side of the
pipe, are particularly significant as they are the missing family of
rotationally-asymmetric waves not found in \cite[]{faisst03,wedin04}
and have the structure preferred by the linear transient growth
mechanism \cite{schmid94}.  They bifurcate from a new mirror-symmetric
family which can be traced down to a saddle node bifurcation at
$Re=773$. This figure substantially lowers the current best estimate
of $1251$ for $Re_g$ - the Reynolds number at which the laminar state
stops being a global attractor.  The relative sizes of this new $Re_g$
and $Re_t$ for pipe flow are then more in line with plane Couette flow
($Re_t = 323$ \cite{bottinc98} and $Re_g =127.7$
\cite{nagata90,waleffe03}: $Re$ based on half the velocity difference
and half the channel width) than plane Poiseuille flow ($Re_t \approx
1300$ \cite{davies28,kao70,patel69,orszag80,carlson82} and $Re_g=860$
\cite{waleffe03}: $Re$ based on the mean flow rate and the channel
width). Beyond suggesting that $Re_g$ in plane Poiseuille flow can be
significantly lowered, these latest discoveries highlight the
substantial delay in $Re$ between new solutions appearing in phase
space and the emergence of sustained turbulent trajectories.
%
% FIG 1
%
\begin{figure}
   \epsfig{figure=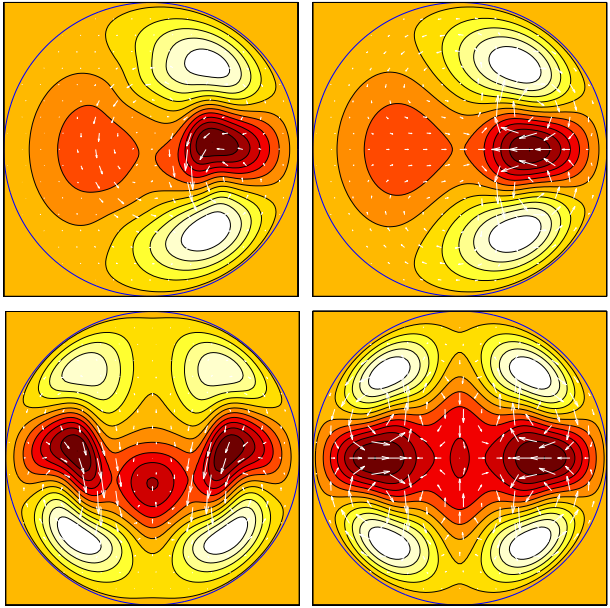, angle=0, width=85mm}
   \caption{\label{fig:slices} Velocity fields for the asymmetric mode 
     at $Re=2900$ (top) and the mirror-symmetric mode at $Re=1344$
     (bottom) (both at $\alpha=0.75$).  An instantaneous state is
     shown on the left and a streamwise-averaged state on the right.
     The coloring indicates the downstream velocity relative to the
     parabolic laminar profile: red(dark) through white(light)
     represents slow through fast (with zero corresponding to the
     shading outside the pipe).  In-plane velocity components are
     shown by vectors. The maximum and minimum streamwise velocities
     (with the laminar flow subtracted) and maximum in-plane speed
     for the asymmetric mode are $0.33$, $-0.42$ and $0.03$
     respectively while for the mirror-symmetric mode they are $0.31$,
     $-0.43$ and $0.08$ (all in units of $U$).}
\end{figure}

%
% FIG 2
%
\begin{figure}
   \epsfig{figure=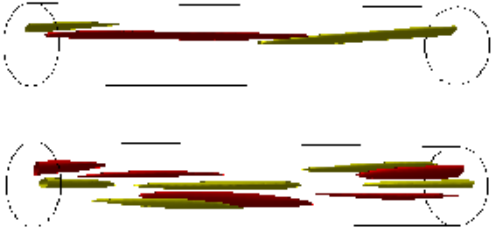, angle=0, width=75mm}
   \caption{\label{fig:vorticity} 
     Instantaneous axial vorticity along one wavelength of the pipe
     for the asymmetric mode (top) and the mirror-symmetric mode
     (bottom) at the same $Re$ and $\alpha$ as Fig \ref{fig:slices}.
     Contours are at +/-60\% of the maximum absolute value
     (green-light/red-dark).  }
\end{figure}
%
% FIG 3
%
\begin{figure}
   \epsfig{figure=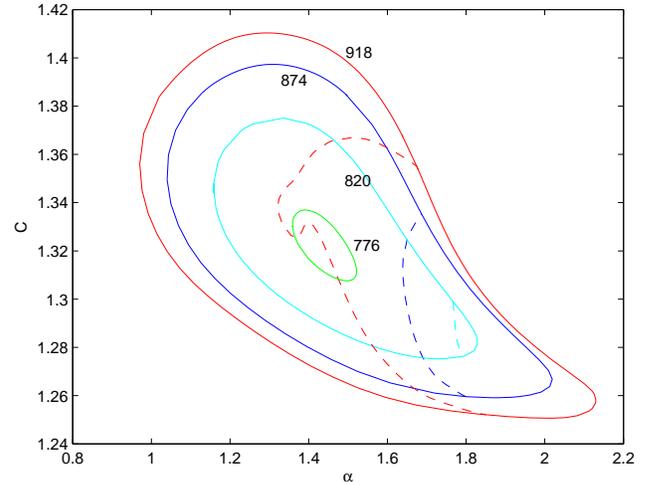, angle=0, width=85mm}
   \caption{\label{fig:alpha} Phase velocity $C$ in units of $U$  as a
   function of $\alpha$ for the mirror-symmetric modes (solid lines)
   and asymmetric modes (dashed) at 4 values of $Re$ near the saddle
   node bifurcation at $Re=773$.  }
\end{figure}

% formulation
The new solutions were captured by inserting a fully 3-dimensional
spectral representation (Chebychev in $s$, Fourier in $\phi$ and $z$
where $(s,\phi,z)$ are the usual cylindrical coordinates aligned with
the pipe) of the velocity and pressure field into the governing
Navier-Stokes equations as viewed from an appropriately rotating and
translating reference frame in which the TW is steady \cite{wedin04}.
The resultant nonlinear algebraic system was solved using the
Newton-Raphson algorithm \footnote{In the nomenclature of
  \cite{wedin04}, typical resolutions used to represent the modes were
  $(15,25,5)$ representing about 20,000 degrees of freedom.}. To start
the procedure off, an artificial body force was added to the
Navier-Stokes equations (see \cite{wedin04}) designed to give
streamwise-independent vortices and streaks of a finite amplitude. The
size of the forcing was then adjusted to find a bifurcation point at
which the translational flow symmetry along the pipe is broken. New
finite-amplitude solutions to pipe flow were found if this solution
branch could be continued back to the zero-forcing limit.

\begin{figure}
   \epsfig{figure=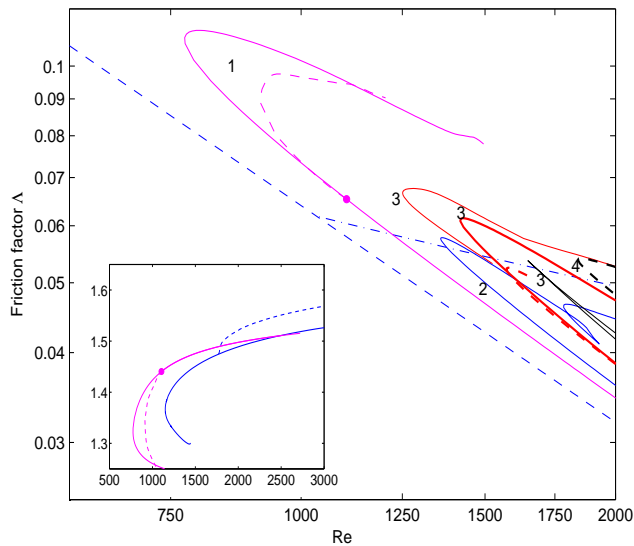, angle=0, width=85mm}
   \caption{\label{fig:friction}
     Friction factor $\Lambda:= 2D \,G/\rho U^2$ against $Re$ for the
     various families of travelling waves where $G$ is the mean
     pressure gradient along the pipe and $\rho$ the density. The
     lower dashed line indicates the laminar value
     $\Lambda_{lam}=64/Re$ and the upper dash-dot line indicates the
     log-law parametrization of experimental data
     $1/\sqrt{\Lambda}=2.0\log(Re \sqrt{\Lambda})$. The labels are $m$
     values for the rotational symmetry $\R_m$ of the different TW
     families all drawn at the wavenumber which leads to the lowest
     saddle node bifurcation. The new TWs shown - mirror-symmetric
     modes (solid) and asymmetric modes (dashed) - correspond to $m=1$
     and $\alpha=1.44$. The bifurcation point is marked with a dot.
     The inset shows the phase velocity $C$ (in units of $U$) versus
     $Re$ for the two types of mode at $\alpha=0.75$ where the
     bifurcation was originally found and the optimum $\alpha=1.44$. }
\end{figure}

The TWs previously isolated \cite{faisst03,wedin04}
were induced using a forcing that was rotationally symmetric under
\[ \R_m: \, (u,v,w,p)(s,\phi,z) \rightarrow  (u,v,w,p)(s,\phi+2 \pi/m,z) \]
for some $m=2,3,4,5$ or $6$.  As well as this rotational symmetry, all
the TWs also possess the shift-\&-reflect symmetry
\[ \S: \, (u,v,w,p)(s,\phi,z)  \rightarrow  (u,-v,w,p)(s,-\phi,z+\pi/\alpha) \]
where $\alpha$ is the base axial wavenumber (so the periodic pipe is
$2\pi/\alpha$ long) and take the form
$\bu(s,\phi,z,t)=\bu(s,\phi,z-Ct)$ where $C$ is the {\em a priori}
unknown axial phase speed of the wave. In contrast, new
rotationally-asymmetric TWs were found by using a forcing function
which created vortices with the radial velocity structure
$u(s,\phi) 
\propto
\Re e\{\, e^{-1/s}(1-s^2)\sum^7_{m=1} [\,1+\cos({\textstyle {m
    \pi\over 7}})\,] e^{im \phi}\, \}$ 
and hence distributed energy across a band of azimuthal wavenumbers.
This choice led to a branch of asymmetric solutions whose component
fast and slow streaks are preferentially located to one side of the
pipe (see Figs \ref{fig:slices} and \ref{fig:vorticity}). These
asymmetric TWs are $\S$-symmetric and have one phase speed $C$ along
the pipe. They extend beyond $Re=5000$ and originate at a pitchfork
bifurcation (at $Re=1770$ when $\alpha=0.75$) from a mirror-symmetric
TW family (see Figs \ref{fig:slices} and \ref{fig:vorticity}) which
satisfies the additional shift-\&-rotate symmetry
\[ 
\O: \, (u,v,w,p)(s,\phi,z)  \rightarrow  (u,v,w,p)(s,\phi+\pi,z+\pi/\alpha)
\]
(coupled with the $\S$-symmetry, this implies invariance
under reflection in the line $\phi=\pm\pi/2$).  The mirror-symmetric
solutions undergo a saddle node bifurcation at much lower $Re$:
$Re=1167$ at $\alpha=0.75$ going down to a minimum of $Re=773$ at
$\alpha=1.44$: see Fig. \ref{fig:alpha}. The friction factors
associated with the upper branches of these new modes are much larger
than the rotationally-symmetric modes \cite{faisst03,wedin04} (see
Fig.  \ref{fig:friction}).

%
% FIG 4
%

%
% FIG 5
%
\begin{figure}
   \epsfig{figure=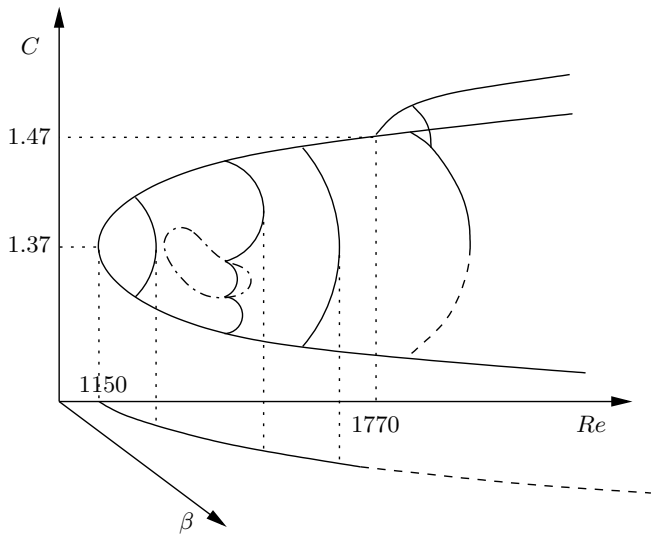, angle=0, width=80mm}
   \put(-180,0){$\beta$}
   \put(-30,37){$Re$}
   \put(-240,180){$C$}
   \put(-245,146){$1.47$} 
   \put(-245,105){$1.37$}  
   \put(-115,37){$1770$} 
   \put(-218,52){$1150$} 
   \caption{\label{fig:bigPicture} A schematic picture of how all the
     new travelling wave branches fit together in $(\beta,Re,C)$ space
     (at $\alpha=0.75$). The main parabolic curve in the $\beta=0$
     plane is the mirror-symmetric branch off which the asymmetric
     branch bifurcates (uppermost line). Helical branches bulge out of
     the $\beta=0$ plane and connect upper and lower parts of the
     mirror-symmetric.  Across a finite range of $Re$, these helical
     modes perforate the $\beta=0$ plane in between the
     mirror-symmetric branches creating an isola of non-helical
     rotating TWs (closed dash-dot loop).  Helical waves also connect the
     asymmetric branch and the helical solutions which originate from
     the mirror-symmetric solutions. }
\end{figure}

\begin{figure}
   \epsfig{figure=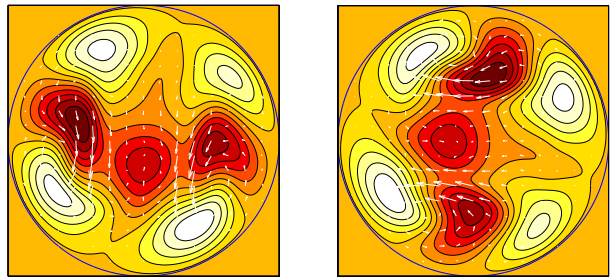, angle=0, width=80mm}
   \caption{\label{fig:helicity} Two velocity slices across a helical
     mode taken at the same instant of time but $25\,D$ apart with
     $\alpha=0.75$, $\beta=0.019$, $\omega=-0.0011$ at $Re=1344$.  The velocity representation is as in Fig. \ref{fig:slices}.}
\end{figure}

\begin{figure*}
   \epsfig{figure=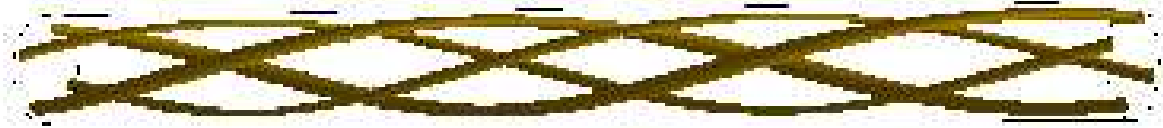, angle=0, width=160mm}
   \caption{\label{fig:faststreaks}
The four fast streaks of the  helical mode shown in Fig. \ref{fig:helicity}
plotted over one $\beta$ wavelength  $\approx 170\,D$.
}
\end{figure*}

Both new families possess the characteristic features of the TWs found
in \cite {faisst03,wedin04}: essentially 2-dimensional fast streaks
near the wall and aligned with the flow, slow streaks in the interior
which are aligned on average with the flow, and a smaller (typically
by an order of magnitude) 3-dimensional wave field.  By continuity,
helical TWs should exist with these fast streaks inclined to the flow
direction and indeed a surface of such solutions can be found
connecting the upper and lower branches of the mirror-symmetric TWs
(see Fig. \ref{fig:bigPicture}).  These helical TWs take the form
$\bu(s,\phi,z,t)=\bu(s,\phi-\beta [z-Ct]-\omega t,z-Ct)$ with $\beta$
measuring the helicity in the Galilean frame moving at $C \bz$ and
$\omega$ being an azimuthal phase speed {\em relative} to the Galilean
frame. Helicity destroys $\S$-symmetry but a modified form of
$\O$-symmetry ($\O_{\beta}$) is preserved where the rotation
transformation is now $\phi \rightarrow \phi+(1-{\beta \over
  \alpha})\pi$.  The helicity $\beta$ and rotational speed $\omega$
never rise above $O(10^{-2})$ for $Re \leq 1500$ confirming the flow
preference for non-rotating, axially-aligned streaks.  Interestingly,
in the range $Re=1165-1330$, the helicity $\beta$ on this surface
passes through zero twice in going between the two mirror-symmetric
branches (see Fig. \ref{fig:bigPicture}). These points correspond to
an isola in the (fixed $\alpha$) $C$ vs $Re$ plane of rotating
non-helical modes which are neither shift-\&-reflect symmetric nor
have any rotational symmetry.  The helical and non-helical rotating
waves look very similar to the mirror-symmetric modes except for a
slight twist in the streak structure along the pipe (see Figs
\ref{fig:helicity} and \ref{fig:faststreaks}). Helical modes continued
off the asymmetric modes have no symmetry at all and originate 
in a symmetry-breaking bifurcation off the $\O_{\beta}$-symmetric
helical solutions extended from the mirror-symmetric waves: see Fig.
\ref{fig:bigPicture}.

%
% FIG 6
%

%
% FIG 7
%

The asymmetric, mirror-symmetric and helical TWs all represent saddle
points in phase space with a very low-dimensional unstable manifolds
(e.g. 2 for the asymmetric mode at $(\alpha,Re)=(0.75,1820)$ and 4 for
the mirror-symmetric mode at $(\alpha,Re)=(0.75,1184)$).  Their
presence indicates the richness of phase space even at Reynolds
numbers approaching $773$.  The delay of transition until $Re \geq
1750$ suggests that the establishment of a `turbulence-bearing'
scaffold constituted of all their stable and unstable manifolds is far
from immediate. The clear implication is that while the emergence of
alternative solutions to the laminar state seems a necessary precursor
for transition, it is {\em not} a good predictor of the actual
Reynolds number at which this occurs in pipe flow (and other shear
flow systems). Once the transitional regime has been reached however,
there is now mounting experimental \cite{hof04,hof05} and numerical
evidence \cite{kerswell07,schneider07a} indicating that some of the
travelling waves at least (those with low to intermediate wall shear
stresses \cite{kerswell07}) appear as transient but recurrent coherent
structures within the flow. Intriguingly, numerical simulations
\cite{kerswell07} have also revealed that a number of travelling waves
with low wall shear stress sit on a dividing surface in phase space (a
separatrix if the turbulence is a sustained state) which separates
initial conditions which directly relaminarise and those which lead to
a turbulent episode. Recent computations \cite{schneider06,eckhardt07}
using a shooting technique to converge onto this dividing surface
appear to have already found that the asymmetric wave sits there too
(compare Fig \ref{fig:slices} to Fig. 8 of \cite{eckhardt07} and Fig.
1 of \cite{schneider06}). The fact that this wave bears some
resemblence to $m=1$ `optimal' disturbances which emerge from linear
transient growth analyses also suggests an enticing opportunity to
bridge the gap between linear and nonlinear approaches.

In summary, we have presented a series of new travelling wave
solutions to the pipe flow problem which have different structure to
existing solutions and which exist at far lower Reynolds numbers.  One
type - the asymmetric modes - represents the missing $m=1$ family from
the waves found initially \cite{faisst03,wedin04}. These waves also
appear to rationalise some interesting results from recent numerical
computations \cite{schneider06,eckhardt07}, thereby corroborating the
picture which is emerging that lower branch TWs (and therefore also
their stable manifolds) sit on the separatrix between laminar and
turbulent states.

\begin{acknowledgments}
  We acknowledge encouraging discussions with Fabian Waleffe
  and the support of EPSRC.
\end{acknowledgments}

\end{document}